\documentclass[preprint]{aastex}

\input{psfig}

\def\lax    {${_<\atop^{\sim}}$}
\def\gax    {${_>\atop^{\sim}}$}

\def\etal   {{\it et al.}~}

\shorttitle{}
\shortauthors{Mathur, Kuraszkiewicz \& Czerny}

\begin{document}

\title{Evolution of Active Galaxies: \\
Black-hole Mass -- Bulge  relations for Narrow Line Objects}


\author{Smita Mathur\altaffilmark{1}}
\affil{Astronomy Department, The Ohio State University, Columbus, OH 43210}
\email{smita@astronomy.ohio-state.edu}
\author{Joanna Kuraszkiewicz}
\affil{Harvard-Smithsonian Center for Astrophysics, Cambridge, MA 02138}
\email{joasia@head-cfa.harvard.edu}
\and
\author{Bozena Czerny}
\affil{N. Copernicus Astronomical Center, Warsaw, Poland}
\email{bcz@camk.edu.pl}


\altaffiltext{1}{Harvard-Smithsonian Center for Astrophysics, Cambridge, MA 
02138}


\begin{abstract}
Mathur (2000) has proposed that the narrow line Seyfert 1 galaxies
(NLS1s) are likely to be the active galaxies in early stage of
evolution. To test this hypothesis, we have calculated the black hole
(BH) masses and the host galaxy bulge masses for a sample of NLS1s. We
find that the mean BH to bulge mass ratio of NLS1s is significantly
smaller than that for normal Seyfert galaxies. We also find that the
ratio of BH mass to bulge velocity dispersion is also significantly
smaller for NLS1s. A scenario of BH growth is our preferred
interpretation, though alternative explanations are discussed.
Assuming that the BHs grow with accretion with a radiative efficiency
of 0.1, it will take them t\gax$3.3\times 10^8$ years to become as
massive as in normal Seyfert 1s. These timescales are consistent with
the theoretical estimates of quasar timescale $t_Q$\lax$4.5\times
10^8$ years calculated by Haehnelt, Natarajan \& Rees (1998).  Studies
of low redshift NLS1s thus provide a powerful, and due to their
proximity relatively easy, way to understand the high redshift quasars
and their evolution.
\end{abstract}




\section{Introduction}

 Last couple of years have seen significant progress in our
understanding of the relationship between quasar activity and galaxy
formation. In the framework of the hierarchical dark-matter cosmogony
(e.g. Haehnelt, Natarajan \& Rees 1998, HNR98 hereafter), the
formation and evolution of galaxies and their active nuclei is
intimately related (Fabian 1999, Granato \etal 1999, Haehnelt \&
Kauffmann 1999). It is assumed that the supermassive black holes form
proportionally to the dark matter halos. The process
of formation of a massive BH and the active nucleus is the very
process of galaxy formation. The active nucleus and the galaxy evolve
together, with BH accreting matter and the galaxy making stars. At one
stage the winds from the active nucleus blow away the matter
surrounding it and a quasar emerges. This is not only the end of
active evolution of the quasar but that of the galaxy as well, which
is evacuated of its interstellar medium. The quasar then shines as
long as there is fuel in the accretion disk (Fabian 1999). In this
scenario, high-redshift quasars represent the early stage of galaxy
evolution, BALQSOs at z$\approx 2$ represent the stage when the gas is
being blown away, and z$\approx 1$ quasars would be the passively
evolving population. Massive ellipticals found today might be the dead
remnants of what were once luminous quasars. 

 Observational support to the above scheme of evolution is also
 accumulating. In a classic paper, Maggorian \etal (1998) found that
 the mass of a BH in the nucleus of a galaxy is correlated with the
 mass of its bulge (see also van der Marel \etal 1999 and an excellent
 review by Richstone \etal 1998). Most recently, Gebhardt \etal (2000a)
 and Ferrarese \& Merritt (2000) found a correlation  between
 the BH mass and the bulge velocity dispersion, a much tighter relation
 than that between the BH mass and bulge luminosity. The important
 point to note is that the nuclear BH seems to ``know'' the galaxy it
 harbors. Laor (1998) showed that the above correlation for normal
 galaxies extends to quasar host galaxies as well. The similarity of
 the two correlations is remarkable given the widely different levels
 of BH activity in luminous quasars and nearby massive normal
 galaxies. Franceschini \etal (1999) found an interesting similarity
 between the evolution rates for the total populations of galaxies and
 AGNs. By comparing the evolution of star formation rate per comoving
 volume in galaxies and the volume emissivity by AGNs, they found that
 the evolution of luminous quasars parallels that of stellar
 populations in massive spheroidal galaxies.

 Much of the above observational evidence relates BH masses of
luminous quasars, or what were once luminous quasars, with spheroidal
masses of massive galaxies. Wandel (1999) extended Laor's work to low
redshift Seyfert galaxies for which BH masses were available using
reverberation mapping. Interestingly, Wandel (1999) found that the
average BH- to bulge-mass ratio for Seyferts is systematically lower
than that for normal galaxies and luminous quasars. Wandel interpreted
this result as a scenario of BH growth in which a BH in a Seyfert
grows by accretion to become a more massive BH in a quasar. This,
however, is highly unlikely since the Seyferts are known to accrete at
a sub-Eddington rate while the luminous quasars in Laor's sample
accrete at close to Eddington rate. So, in the above scenario, the
accretion rate has to grow with time, which is difficult to achieve
and contrary to theoretical expectations. A possible explanation of
the Laor and Wandel results might be that the quasars and Seyferts
represent two different populations (Mathur 2000). The quasar
phenomenon may be a result of galaxy formation due to primordial
density fluctuations or mergers at early cosmic epochs, which were to
result in massive ellipticals of today. At low redshifts, when new
galaxies are formed due to interactions or mergers, similar evolution
may take place. Gebhardt \etal (2000b) have now found that the
Seyferts and normal galaxies have the same BH mass--bulge velocity
dispersion relationship. Note, however, that it still implies that the
Seyferts are different from the normal galaxies in that the bulges of
Seyferts have lower mass to light ratio, unless the Wandel result is
purely due to observational uncertainty.

While we now have the observational evidence for a relation between an
AGN and its host galaxy, we do not know how it got there, or what are
the evolutionary steps towards the ``final'' correlation discussed
above.  Mathur (2000) has argued that narrow line Seyfert 1 galaxies
(NLS1s) may represent an early phase in the evolution of active
galaxies at low redshift.  In our scenario, the luminous quasars and
the low-luminosity Seyferts follow separate but similar evolutionary
tracks. In both cases the accretion rate \.m = \.M$/$\.M$_{Eddington}$
is large in the early stages of evolution and reduces later on,
consistent with the theoretical expectations (e.g. HNR98). To test
this hypothesis, we determine the BH- to bulge-mass relation as well
as the BH-mass to bulge velocity dispersion relation for NLS1s and
compare those with the normal Seyfert relation. In $\S 2$ we detail
our method and present interpretation in $\S 3$.

\section{BH and Bulge  Relations for NLS1s}

\subsection{Determination of BH Mass}


Our sample consists of 15 narrow-line AGN (11 NLS1s and 4 NL quasars)
for which soft- and hard X-ray spectra are available in literature.
The central black hole masses for these objects were calculated by
fitting their spectral energy distributions with the accretion disk
and corona (ADC) model of Witt, Czerny \& \.Zycki (1997) (see
Kuraszkiewicz \etal~2000 for details).  In this model the corona
accretes and generates energy through viscosity, and the division
between optically thin accretion flow (corona) and optically thick
flow (disk) results from the cooling instability discussed by Krolik,
McKee and Tarter (1981). The model is fully described by the following
parameters: the mass of the central black hole, accretion rate and
viscosity parameter.  The calculated black hole masses are quoted in
Table~1.

We  determine the errors of the black hole mass estimated from
spectral fitting as follows. The fit is based on 
 two relations:  
(1) the asymptotic shape of the disk emission
in the optical band is: $F_{\nu} \propto \nu^{1/3} (M \dot M)^{2/3}
cos i H_0^2 f^{-4/3}$, where $M$ is mass of the black hole, $\dot M$ -
accretion rate, $i$ - the inclination angle of the disk, $H_0$ is the
Hubble constant, f - spectral correction given by color temperature to
the effective temperature ratio; 
(2) the bolometric luminosity is:
$F_{bol} \propto \dot M \eta H_0^2$, where $\eta$ is the efficiency of
accretion.  
In our fits we adopt the following values: $H_0=50$~km/s/Mpc, $cos
i=1$, $f=1$ and $\eta=1/16$. If any of these values are
different, the mass of the black hole would scale as:
$M_{real}=M_{calc} f^2 (cos i)^{-1}(H_0/50)^{-1}$. The inclination
angle of the disk for Seyfert~1s is in the range $0^o$ to $48^o$
(Schmitt et al. 2000 and assuming the accretion disk is aligned with
the dusty torus), the upper limit for f is 1.7 (R\'o\.za\'nska,
private communication), and if there is strong outflow close to the
black hole $\eta$ decreases (up to a factor 2 if
the outflow is energetically comparable to the outflows in radio-loud
objects).  Taking all these effects into account we estimate the
calculated mass of the black hole to be accurate to $^{+0.64}_{-0.30}$
in logarithm.

For the sake of consistency and comparison with earlier results, we
also calculated the BH masses for the 7 Sy1s from Wandel's sample, for
which X-ray spectra were available. The results are presented in
Table~2.  The calculated black hole masses are compared to the virial
masses in Tables~2 and 3. The masses calculated by the ADC model are
generally larger than masses estimated by Wandel.  We
calibrate our BH mass determinations with the virial mass estimates
for the objects in Table~2 and 3 and find that:
$\log M_{\rm bh}^{\rm calib} = \log M_{\rm bh}^{\rm calc}
-0.70\pm0.15$.  This relation, together with the calculated and
virial BH masses for Seyfert 1s and NLS1s (excluding the outlying
point for IC 4329A\footnote{The virial mass for IC 4329A is adopted
from Wandel 1999 who has adopted it from Wandel, Peterson \& Malkan
1999. We find that IC 4329A is an outlier in all the relations
discussed in Wandel, Peterson \& Malkan 1999 as well as in Peterson
\etal 2000. This leads us to suspect the virial BH mass of this
object. We ignore IC 4329A in rest of the paper.}), is displayed in
Figure 1. It can be clearly seen that the above equation is a good
description of the relation between BH masses calculated by the ADC
model and by the virial method. There is clearly some uncertainty in
the ADC mass estimates, as discussed above (Tables 1, 2, 3), and the
virial mass estimates are also uncertain by a factor of several
because of the uncertainty in the geometry of the broad emission line
region. Both of these effects contribute towards the scatter seen in
Figure 1.  The calibrated black hole masses for NLS1s are given in
Table~3. These masses were used to calculate the black hole to bulge
mass ratio.

\subsection{Determination of Bulge Mass}

For the NLS1s in our sample, we determined the bulge masses of their
host galaxies following the general method used by Laor (1998) and Wandel 
(1999),
which we discuss briefly below.  For most of our NLS1s we obtain the
bulge absolute blue magnitudes ($M_{\rm B}^{\rm bulge}$) from Whittle
(1992).
A galaxy's total blue magnitude ($M_{\rm
B}^{\rm total}$) is related to $M_{\rm B}^{\rm bulge}$ by Simien \& 
Vaucouleurs (1986) equation
(for $H_0=50$~km~s$^{-1}$~Mpc$^{-1}$, $q_0=0$):
$ M_{\rm B}^{\rm total} - M_{\rm B}^{\rm bulge} = 0.324\tau - 
0.054\tau^2+0.0047\tau^3$
where $\tau=T+5$ and T is the Hubble stage (defined in de Vaucouleurs, 
de Vaucoulers \& Corwin 1976). 


Mrk~110 and Mrk~335 do not have a Hubble type defined in the Whittle
(1992) compilation. However NED\footnote{This research has made use of
the NASA/IPAC Extragalactic Database (NED) which is operated by the
Jet Propulsion Laboratory, California Institute of Technology, under
contract with the National Aeronautics and Space Administration.}
quotes a S0/a Hubble type for Mrk~335, which we use to calculate the
bulge magnitude for this object. The resulting $M_{\rm B}^{\rm bulge}$
(Table 3) is consistent with that in Ho (1999). For Mrk~110 we adopt a
canonical Hubble type of Sa.

For NAB~0205+024, PG~1402+261 and PG~1444+407 we take the host galaxy
absolute magnitudes from Bahcall et al. (1997), which have been
corrected for nuclear emission. For Mrk 1044 we take the host galaxy
magnitude from MacKenty (1990), who has nuclear emission included in
the total blue magnitude, hence in Table 3 we quote the bulge mass 
as an upper limit.

After obtaining the bulge blue magnitudes we use the relation between
the bulge B and V magnitude (Worthey 1994) B$-$V=0.95, the standard
relation between the bulge luminosity and absolute magnitude: $\log
(L_{\rm bulge}/L_{\odot}) = 0.4(-M_{\rm V}^{\rm bulge}+4.83)$ and the
relation between bulge mass and bulge luminosity for normal galaxies
from Magorrian et al. (1998): $\log M_{\rm bulge}/M_{\odot} =
-1.11+1.18 \log L_{\rm bulge}/L_{\odot}$. The bulge masses calculated
this way are listed in Table~3. 

\subsection{Determination of Bulge Velocity Dispersion}

We do not have bulge velocity dispersion ($\sigma_{\star}$)
measurements for the objects in our sample. However, Nelson (2000) has
shown that the width of the narrow emission line [OIII] serves as a
good representation of the velocity dispersion with $\sigma_{\star}$ =
FWHM [OIII]/2.35. Adopting from Nelson (2000), we collected from
literature, as well as from our optical data, the values of FWHM [OIII] for all the NLS1s in our sample 
(Osterbrock \& Pogge 1985, Wamsteker \etal 1985, Dahari \& De Robertis
1988, Busko \& Steiner 1988, Nelson \& Whittle 1995, Wilkes \etal 1999).


\section{Results and Discussion}


\subsection{BH- to Bulge-mass relation for NLS1s}

In Figure 2 we have plotted the $\log$ of BH mass vrs. $\log$ of bulge
mass obtained using the method discussed in $\S 2$. The dashed line
represents the relation for normal galaxies and luminous quasars (from
Laor 1998) while the dotted line represents the relation for Seyfert
galaxies (from Wandel 1999, excluding 3 NLS1s in his sample; we have
also re-calculated the bulge masses for these objects, as Wandel uses
a slightly different relation between the mass and luminosity of the
bulge than we do). The NLS1s are represented by filled squares.  It is
immediately apparent that the relation for Seyfert galaxies does not
hold for NLS1s; NLS1s lie systematically below the line for Seyfert
galaxies.  In addition, we find that the narrow line quasars also lie
systematically below the line for quasars.
The mean $M_{\rm BH}/M_{\rm bulge}$ for NLS1s is 0.00005, lower by a
 factor of six compared to $M_{\rm BH}/M_{\rm Bulge}=0.0003$ for
 Wandel's Seyferts, excluding the three NLS1s in his
 sample. Similarly, the BH to bulge mass ratio for NL quasars is
 0.0005, almost an order of magnitude lower than 0.006 for normal
 galaxies and quasars.

 We would like to emphasize that, since we have calibrated our BH
 masses with those from the reverberation mapping, the relative
 difference in the BH- to bulge-mass relation of NLS1s and normal Sy1s
 is a robust result, though the absolute values may suffer from
 various uncertainties. We have not been able to do such a calibration
 for NL quasars, so we cannot place similar confidence in the
 difference between NL quasars and the quasars in Laor's sample.

 Comparison of BH mass to bulge mass, using bulge luminosity, is a
worthwhile exercise even in the light of the new, stronger,
correlation with bulge velocity dispersion. Reasons for the
differences in two correlations may be many: e.g. Hubble types of many
bright AGNs may have systematic errors, or nuclear starbursts may
reduce the mass to light ratio of bulges in AGNs (Gebhardt \etal
2000b). Understanding these differences is important. Moreover, here
we are comparing two subclasses of Seyfert galaxies: regular, broad
line Seyfert 1s and NLS1s, so any systematic problems in determining
the bulge luminosity is likely to be the same in both classes. So, any
difference in the two is not likely to be a result of observational
uncertainty.

\subsection{BH-mass to Bulge-velocity dispersion relation for NLS1s}

Nelson (2000) has determined BH-mass to bulge velocity dispersion
relation for Seyfert galaxies. It is noteworthy that all the three
NLS1s in his sample: NGC 4051, Mkn 110, and Mkn 335, lie at the lowest
boundary of the scatter.  In figure 3 we have plotted the $\log$ of BH
mass vrs FWHM [OIII]/2.35. The open squares are the Seyfert 1s and the
filled circles are the NLS1s in our sample. The straight line
represents the relation for Seyfert galaxies from Nelson (2000). While
the Seyfert 1s follow the Nelson relation, NLS1s lie systematically
below the line.

\subsection{Interpretation}

The above results support our conjecture that NLS1s are young AGNs.
 NLS1s are believed to accrete at close to Eddington rate, as first
 proposed by Pounds, Done \& Osborn (1995). As discussed earlier, in
 our scenario the accretion rate is highest in the early growing stage
 of an AGN. The BH would grow by accretion and eventually the
 accretion rate would slow down. Whether the bulge mass would grow or
 not would depend upon how young/ rejuvenated the galaxy is, or how
 complete is the interaction/ merger. So an AGN will follow a
 vertically upward or diagonally upward (with both $M_{\rm BH}$ and
 $M_{\rm Bulge}$ increasing) track on Figure 2, as it
 evolves. Similarly, if the bulge velocity dispersion has settled
 down to its ``final'' value, an AGN would follow a vertically upward
 track on Figure 3.

An alternative interpretation of the BH- to bulge-mass relation might
be that the bulge mass to light ratio in NLS1s is lower compared to
normal Seyferts (see $\S 1$). This might be due to higher bulge
luminosity as a result of a nuclear starburst, again supporting our
proposal of NLS1s being rejuvenated AGN (Mathur 2000). This, however,
cannot explain the result regarding velocity
dispersion. Alternatively, the above results might mean that the bulge
luminosities and FWHM [OIII] are systematically overestimated for
NLS1s. While not impossible, it seems unlikely that both the
quantities are overestimated.

Assuming that our results are not just due to observational
uncertainties, and that our scenario of BH growth is valid, some
interesting inferences may be obtained.  It can be seen from figure 2
that for NLS1s, the relation between BH mass and bulge mass is not
strictly linear. So, the BH mass is not a ``universal fraction'' of
the baryonic mass as suggested by Haiman \& Loeb (1998). More likely,
the accretion process determines the BH mass (e.g. HNR99) at any given
time.  The Salpeter time for the growth of a BH, i.e. the e-folding
time, is $t_s = 4 \times 10^7 (L_{Eddington}/L)\eta_{0.1}$ years where
$\eta_{0.1}$ is the radiative efficiency normalized to 0.1.
$L/L_{Eddington}$ calculated for the objects in our sample is listed
in Table~1.  For a bulge with mass $\approx 10^{11}$M$_{\odot}$, we
have a BH mass of $5\times 10^{6}$M$_{\odot}$ for NLS1
Mrk~493. (Figure 2, Table 3). Let us assume this to be the initial BH
mass. The BH would grow to a ``final'' Seyfert mass of $\sim 5\times
10^7$M$_{\odot}$ assuming $M_{\rm BH}/M_{\rm Bulge}=0.00047$. So to
grow by a factor of 10, a BH would need $2.3t_s=3.3\times 10^8
\eta_{0.1}$ years, for a constant accretion rate. Since the accretion
rate decreases with time, the growth time scale t\gax$3.3\times 10^8
\eta_{0.1}$ years.

It is possible that being more luminous, quasars evolve faster than
the Seyferts. However, if the evolution of quasars and of Seyferts is
indeed similar, then quasar lifetimes of about
$10^8$ years is inferred.  This timescale is much larger than $\sim
10^6$ yr. proposed by Haiman \& Leob (1998). HNR98 predict quasar
lifetimes anywhere in the range $10^6$ to $10^8$ yrs. Our results are
consistent with, and close to their upper limit $t_Q$\lax$4.5\times
10^8 \eta_{0.1}$ years calculated for $L/L_{Eddington}=1.0$. HNR98
have considered three different scenarios for BH growth in a quasar:
(i) accretion far above the Eddington rate, (ii) accretion obscured by
dust, and (iii) accretion below the critical rate leading to an
advection dominated accretion flow lasting for a Hubble time. Our
results, together with Laor (1998) make the third option unlikely. For
NLS1s, we have evidence for accretion rate close to Eddington rate,
but not far above the Eddington rate. There are also several NLS1s
which are dusty (e.g. IRAS~13349+2438). So options (i) and (ii) are
plausible. Studies of low redshift NLS1s thus provide a unique
and relatively easy way to understand the high redshift quasars and
their evolution.

We gratefully acknowledge financial support through NASA grant
NAG5-8913 (LTSA to SM) and NAG5-3363task22 (JK).





\clearpage
{\bf Figure Captions:}\\

\noindent
Figure 1: $log$ of BH masses estimated by reverberation mapping is
plotted against that calculated with ADC model. Filled symbols are for
Seyfert 1s and the open symbols are for NLS1s. The straight line
represents the relation between the two mass estimates (see text).

\noindent
Figure 2: BH mass plotted against bulge mass. The NLS1s, represented by filled
squares lie systematically below the dotted line describing the
Seyfert relation. 

\noindent
Figure 3: BH mass plotted against the bulge velocity
dispersion. Seyfert 1s are represented by open squares and NLS1s by
filled circles. The straingt line describes the AGN relation from
Nelson (2000).



\begin{figure}[h]
\vspace*{-1.0in}\hspace*{-1.0in}
\psfig{file=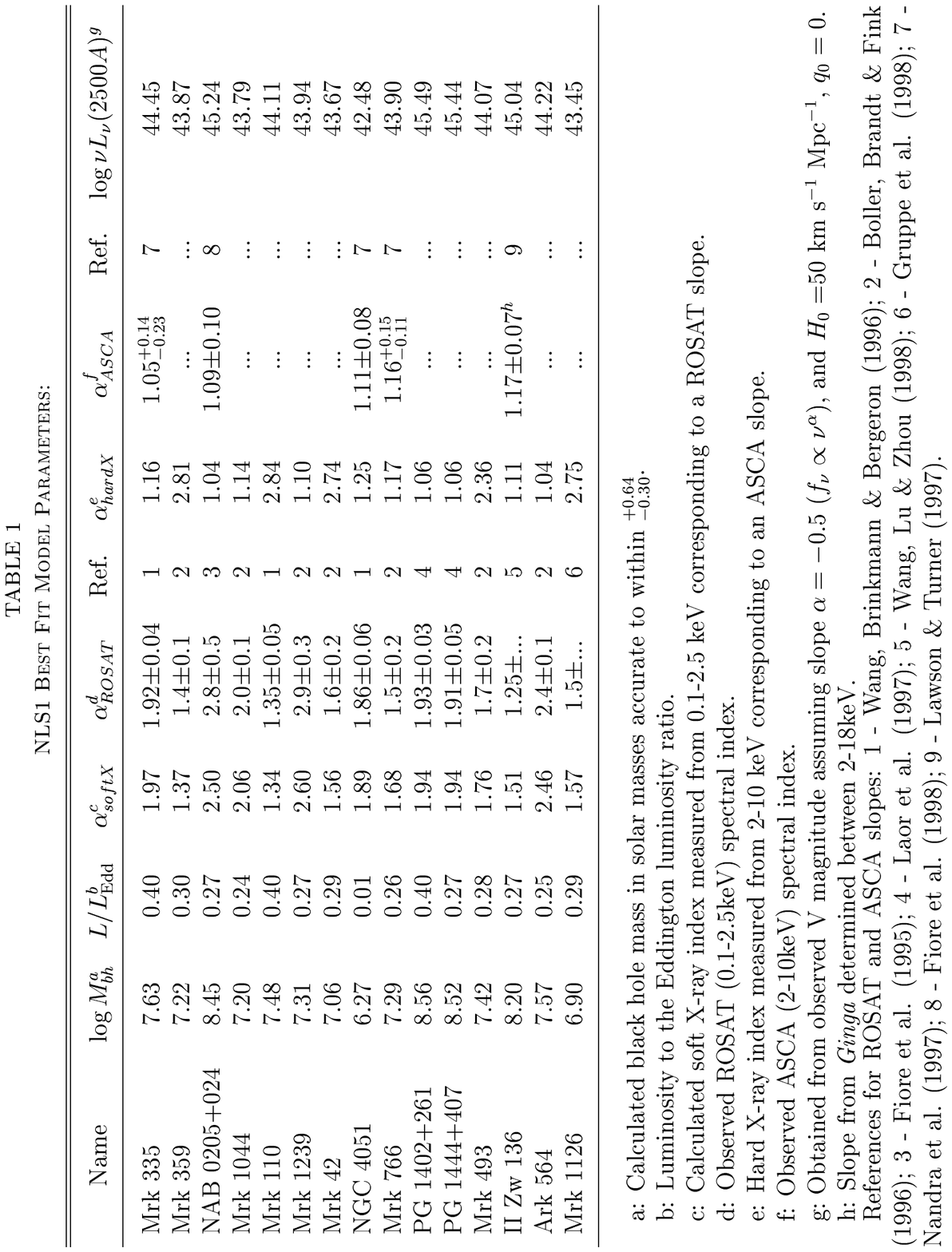,angle=180}
\end{figure}

\begin{figure}[h]
\vspace*{-2.0in}\hspace*{-1.0in}
\psfig{file=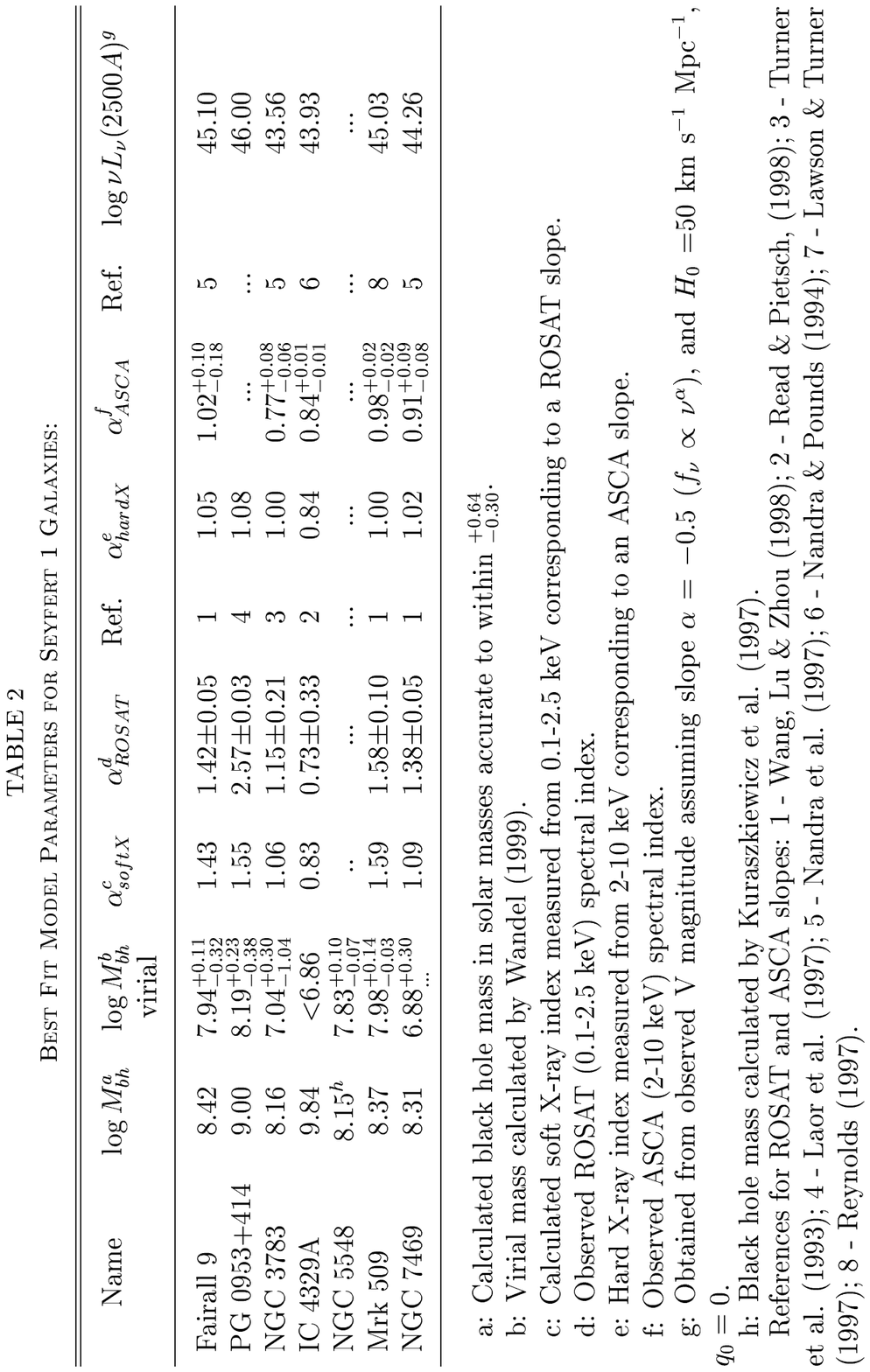,angle=180}
\end{figure}

\begin{figure}[h]
\vspace*{-1.0in}\hspace*{-1.0in}
\psfig{file=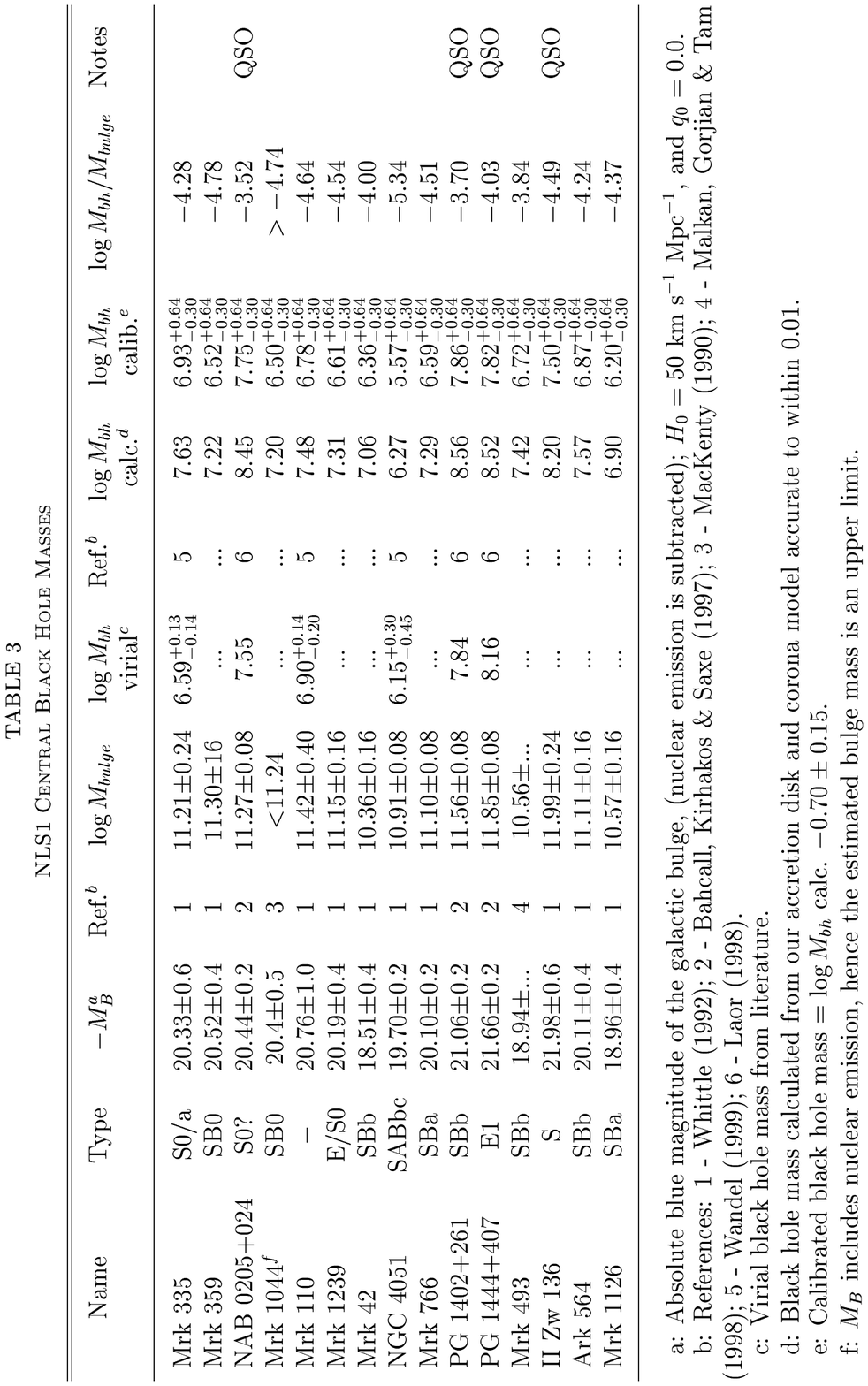,angle=180}
\end{figure}

\begin{figure}[h]
\psfig{file=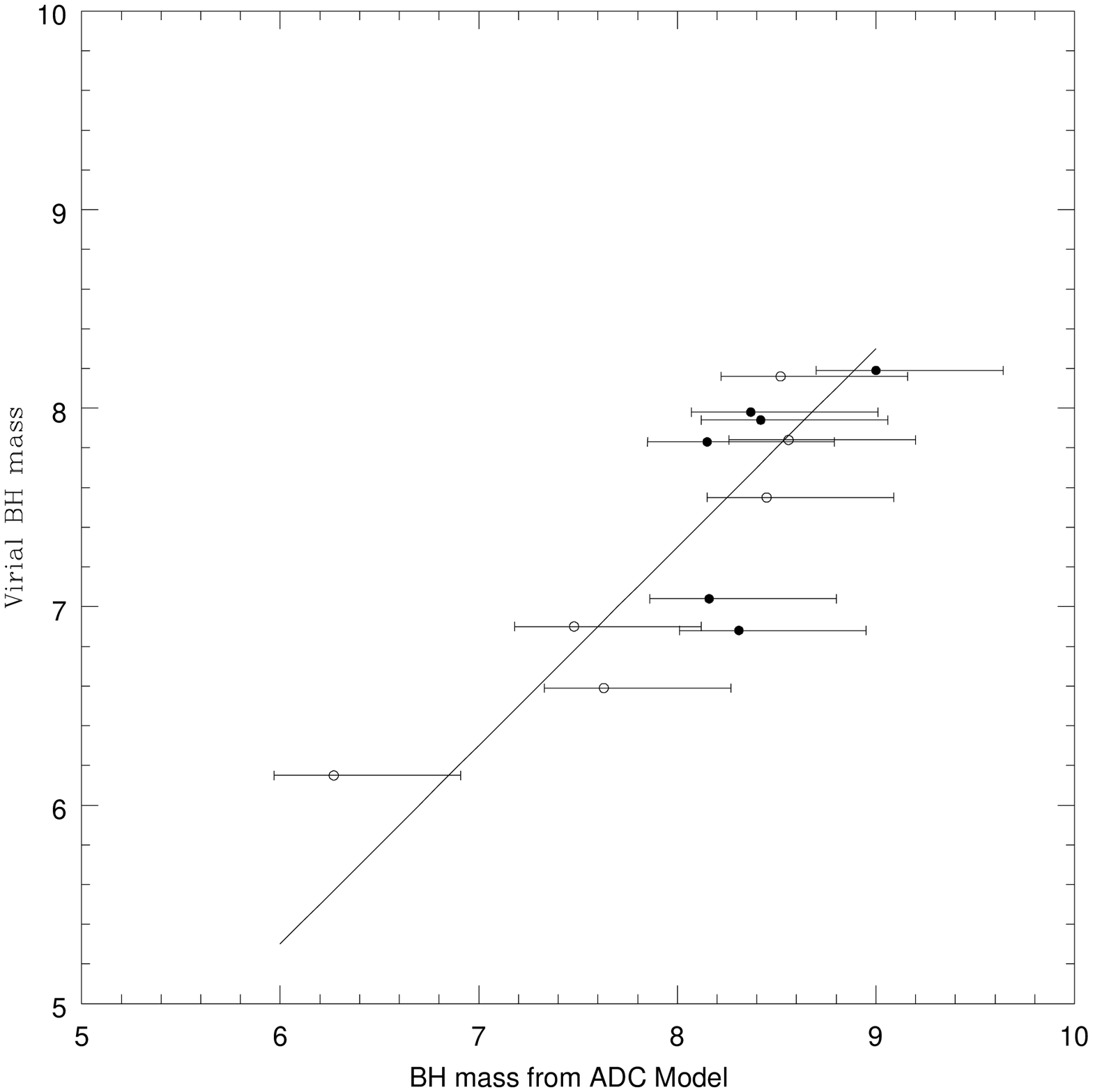,height=6.0truein,width=6.0truein}
\caption{}
\end{figure}

\begin{figure}[h]
\psfig{file=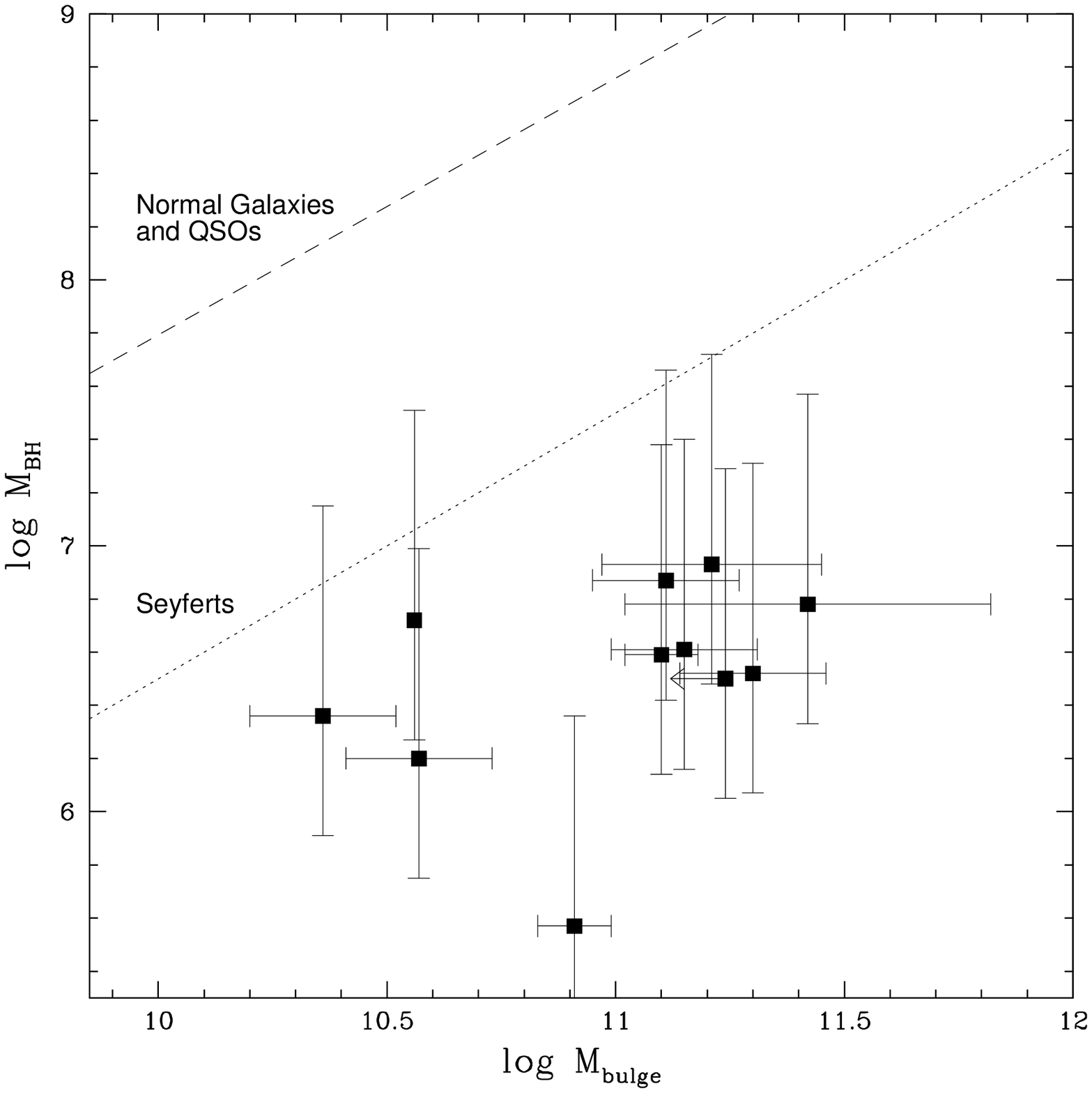,height=6.0truein,width=6.0truein}
\caption{}
\end{figure}

\begin{figure}[h]
\psfig{file=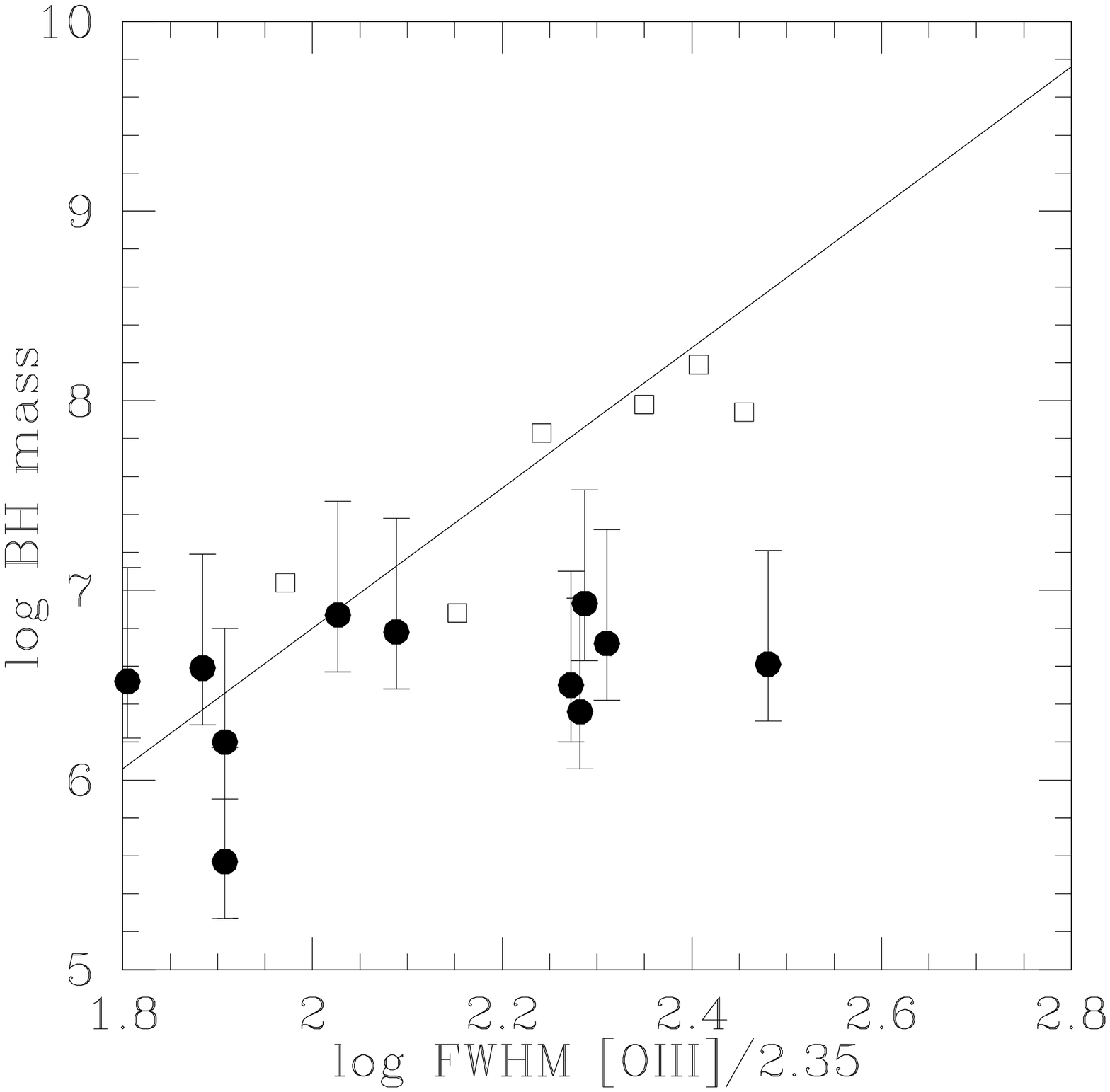,height=6.0truein,width=6.0truein}
\caption{}
\end{figure}


\begin{thebibliography}{}
\bibitem[]{} Bahcall, J. N., Kirhakos, S., Saxe, D. H., \& Schneider,
D. P. 1997, ApJ, 479, 642
\bibitem[]{} Boller, Th., Brandt, W.N., \& Fink, H. 1996, A\&A, 305, 53
\bibitem[]{} Busko, I.C. \& Steiner, J.E. 1988, MNRAS, 232, 525
\bibitem[]{} de Vaucouleurs, G., de Vaucouleurs, A., \& Corwin, H.G.,
Jr. 1976, The Second Reference Catalogue of Bright Galaxies (Austin:
Univ. Texas Press)
\bibitem[]{} Dahari, O. \& De Robertis, M. 1988, ApJS, 67,249
\bibitem[]{} Fabian, A.C. 1999, MNRAS, 308, L39
\bibitem[]{} Ferrarese, L. \& Merritt, D. 2000, ApJ, 539, L9
\bibitem[]{} Fiore, F., Elvis, M., Siemiginowska, A., Wilkes, B. J., 
McDowell, J. C., \& Mathur, S. 1995, ApJ, 449, 74
\bibitem[]{} Fiore, F., Matt, G., Cappi, M., Elvis, M., Leighly,
K. M., Nicastro, F., Piro, L., Siemiginowska, A., \& Wilkes, B. J. 1998,
MNRAS, 298, 103
\bibitem[]{} Franceschini, A., Hasinger, G., Miyaji, T., \& Malquori, D. 1999, 
MNRAS Letters, 310, L5
\bibitem[]{} Gebhardt, K. \etal 2000a, astro-ph/0006289
\bibitem[]{} Gebhardt, K. \etal 2000b, astro-ph/0007123
\bibitem[]{} Granato, G.L., Silva, L., Monaco, P., Panuzzo, P., Salucci, P.,
De Zotti, G., \& Danese, L. 1999 MNRAS, submitted (astro-ph/9911304)
\bibitem[]{} Grupe, D., Wills, Beverley J., Wills, D., \& Beuermann,
K. 1998, AA, 333, 827
\bibitem[]{} Haehnelt, M. \& Kauffmann, G. 2000, MNRAS, 318, L35
\bibitem[]{} Haehnelt, M., Natarajan, P. \& Rees, M. 1998 MNRAS, 300, 817 (HNR98)
\bibitem[]{} Haiman, Z. \& Loeb, A. 1998, ApJ, 503, 505
\bibitem[]{} Ho, L. 1999 in ``Observational Evidence for Black Holes in the Universe'', Ed: S. Chakrabarti (Dordrecht: Kluwer)
\bibitem[]{} Krolik, J. H., McKee, C. F., \& Tarter, C. B. 1981, ApJ, 249, 
422 
\bibitem[]{} Kuraszkiewicz, J., Wilkes, B.J., Czerny, B., \& Mathur, S. 2000, 
ApJ, 542, 692
\bibitem[]{} Laor, A., Fiore, F., Elvis, M., Wilkes, B. J., \&
McDowell J. C. 1997, ApJ, 477, 93
\bibitem[]{} Laor, A. 1998, ApJL, 505, 83
\bibitem[]{} Lawson, A. J.,  \& Turner, M. J. L. 1997, MNRAS, 288, 920
\bibitem[]{} MacKenty, J. W. 1990, ApJS, 72, 231 
\bibitem[]{} Magorrian, J., Tremaine, S., Richstone, D., Bender, R.,
Bower G., Dressler, A., Faber, S. M., Gebhardt, K., Green, R.,
Grillmair, C., Kormendy, J. \& Lauer T. 1998, ApJ, 115, 2285
\bibitem[]{} Malkan, M. A., Gorjian, V., \& Tam, R. 1998, ApJS, 117, 25
\bibitem[]{} Mathur, S. 2000, MNRAS Letters, 314, L17
\bibitem[]{} Nandra, K., \& Pounds, K.A. 1994, MNRAS, 268, 405 
\bibitem[]{} Nandra, K., George, I. M., Mushotzky, R. F., Turner,
T. J., \& Yaqoob, T. 1997, ApJ, 477, 602
\bibitem[]{} Nelson, C. H. \& Whittle, M. 1995, ApJS, 99, 67
\bibitem[]{} Nelson, C. H. 2000, ApJL, 545, 91
\bibitem[]{} Osterbrock, D. \& Pogge, R. 1985, ApJ, 297, 166
\bibitem[]{} Peterson, B.M. \etal 2000, ApJ, 542, 161
\bibitem[]{} Pounds, K., Done, C., \& Osborne, J., 1995 MNRAS, 277, L5
\bibitem[]{} Read, A. M., \&  Pietsch, W., 1998, A\&A, 336, 855
\bibitem[]{} Richstone, D. \etal 1998, astro-ph/9810378
\bibitem[]{} Reynolds, C. S. 1997 , MNRAS, 286, 513
\bibitem[]{} Schmitt, H.R.,  Antonucci, R.R., Ulvestad, J.S., Kinney, A.L., Clarke, C.J., \& Pringle, J.E., 2001, ApJ, in press (astro-ph/0103263) 
\bibitem[]{} Simien, F., \& de Vaucouleurs, G. 1986, ApJ, 302, 564
\bibitem[]{} Turner, T. J., Nandra, K., George, I. M., Fabian, A. C., \&
Pounds, K. A. 1993, ApJ, 419, 127,
\bibitem[]{} van der Marel, R. P. 1999 in `Galaxy Interactions at Low and High 
Redshift', Proceedings of IAU Symposium 186, held at Kyoto, Japan, 26-30 
August, 1997. Ed: J. E. Barnes, and D. B. Sanders. Kluwer Academic Publishers, 
Dordrecht/Boston/London, 1999., p.333
\bibitem[]{} Wamsteker, W. \etal 1985, ApJ, 295, L33
\bibitem[]{} Wandel, A. 1999, ApJL, 519, 39
\bibitem[]{} Wandel, A., Peterson, B.M., \& Malkan, M. 1999, ApJ, 526, 579 
\bibitem[]{} Wang, T.-G., Lu, Y.-J. \& Zhou, Y.-Y. 1998, ApJ, 483, 1
\bibitem[]{} Wang, T., Brinkmann W., \& Bergeron J. 1996, A\&A, 309, 81
\bibitem[]{} Whittle, M. 1992, ApJS, 79, 49
\bibitem[]{} Wilkes, B.J. \etal, 1999, ApJ, 513, 76  
\bibitem[]{} Witt, H. J. Czerny, B., \& \.Zycki, P. T. 1997, MNRAS, 286, 848
\bibitem[]{} Worthey, G. 1994, ApJS, 95, 107
\bibitem[]{}
\end{thebibliography}
\end{document}